\documentclass[letterpaper,12pt]{article}

\usepackage[english]{babel}
\usepackage[applemac]{inputenc}
\usepackage{graphicx} 
\usepackage{booktabs} 
\usepackage{amsmath}  
\usepackage{amssymb}
\usepackage{amsthm}
\usepackage[colorlinks=true,linkcolor=blue]{hyperref} 
\usepackage{memhfixc} 
\usepackage{makeidx} 
\usepackage[font=scriptsize]{caption}
\usepackage[font=scriptsize]{subcaption}  
\usepackage{tikz}
\usetikzlibrary{arrows}
\usetikzlibrary{automata}

\numberwithin{equation}{section}
\numberwithin{figure}{section}

\theoremstyle{definition} 
\theoremstyle{definition} 
\theoremstyle{definition} 
\theoremstyle{definition} 
\theoremstyle{definition} 

\author{Alejandro Nieto Ramos, Elizabeth M. Cherry}
\title{Efficient Representations of Cardiac Spatial Heterogeneity in Computational Models}
\date{Rochester Institute of Technology}

\begin{document}

\maketitle

\section{Introduction}

There is a growing need for quantitative accurate models for use in applications from personalized health care to drug development. Although it is possible to identify parameter values that allow a model to match specific experimental data, most models neglect the intrinsic differences in cardiac cells that result in spatial heterogeneity of electrophysiological properties. These differences are not well described but can be observed using techniques like optical mapping. Nevertheless, developing individualized models for each cell would be prohibitively expensive. In this study, we examine the use of coarse grids for efficiently representing spatially varying parameter values in cardiac tissue models. 

\section{Methods}
To assess how to incorporate observed heterogeneity into cardiac models efficiently, we use the Fenton-Karma model with heterogeneity included as a smooth gradient over space for one or more model parameters. The Fenton-Karma model is a system of differential equations that uses the minimum set of ionic membrane currents to characterize action potentials. 

\begin{align*}
\frac{\partial u(t)}{\partial t}&= D \frac{\partial^2 u}{\partial x^2}-I_{ion}(u,v,w),\\
\frac{\partial v(t)}{\partial t}&= \left \{ 
			\begin{array}{rl}
  				-\frac{v}{\tau_v^+}, & u\ge u_c\\
				\frac{1-v}{\tau_{v1}^-}, & u_c > u \ge u_v\\
				\frac{1-v}{\tau_{v2}^-}, & u<u_v,\\
 			\end{array} \right.\\
\frac{\partial w(t)}{\partial t}&= \left \{
			\begin{array}{rl}
  				-\frac{w}{\tau_w^+}, & u\ge u_c\\
				\frac{1-w}{\tau_{w}^-}, & u<u_c,\\
 			\end{array} \right.			
\end{align*}
where
\begin{align*}
I_{fi}&= \left \{ 
			\begin{array}{rl}
  				-\frac{(1-u)(u-u_c)v}{\tau_d}, & u\ge u_c\\
				0, & u<u_c\\
 			\end{array} \right.\\
I_{so}&= \left \{
			\begin{array}{rl}
  				-\frac{1}{\tau_r}, & u\ge u_c\\
				\frac{u}{\tau_0}, & u<u_c,\\
 			\end{array} \right.\\
I_{si}&=-\frac{w}{2\tau_{si}}(1+\tanh(k(u-u_c^{si}))).			
\end{align*}

The term $D \frac{\partial^2 u}{\partial x^2}$ is a diffusion term and $I_{ion}$ is the sum of the fast inward current $I_{fi}$, the slow outward current $I_{so}$, and the slow inward current $I_{si}$, which basically represent the sodium, potassium, and calcium currents, respectively. 

The simulation is started with a known spatial variation in a model parameter value. Four mathematical nonsymmetric functions are used to represent how the parameter varies in space, including a sigmoid, a quadratic, a cubic, and a sinusoid of varying magnitude. Each shape is considered in each direction along the cable (e.g., both an original sigmoid shape and another reflected). 
Eight different parameters from the model that play substantial roles in determining dynamical properties are studied: $\tau_v^+, \tau_{v1}^-, \tau_w^+, \tau_w^-, \tau_d, \tau_r, \tau_{si}$ and $k$. Parameter values are varied over ranges that allow significant differences in dynamics. 

The model equations are solved on a uniform grid with a spacing of 0.025 cm and a time step of 0.1 ms using forward differences in time and central differences in space. The simulations are made for 10 seconds and the initial conditions were the same for almost all cases. We studied cable lengths of 12.8 and 25.6 cm. 

\begin{figure}[h]
	\centering
	\includegraphics[angle=0, width=0.9\textwidth]{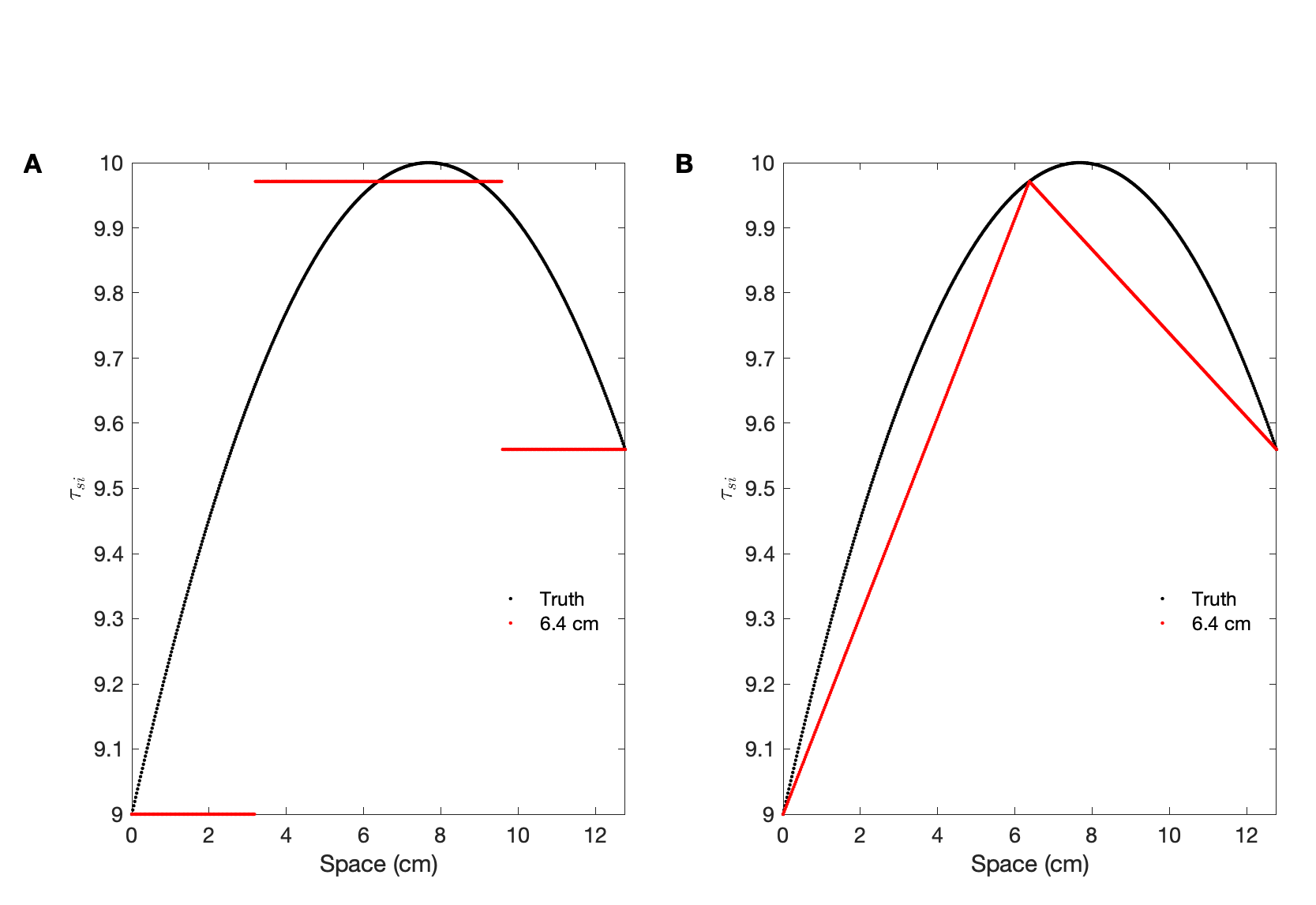}
	\caption{\label{approx}Example of the two kinds of the approximation using a quadratic original function for the parameter $k$ for a cable of 12.8 cm. In graph A we see the piecewise (pw) constant approximation and in graph B the piecewise linear. The coarser grid for a cable this length is used in the example.}
\end{figure}

To assign parameter values, first a uniform parameter grid with specified spacing was developed. Parameter values were assumed to be known exactly at the grid points of the parameter grid, which coincided with a subset of points on the computational grid. Then parameter values were assigned to the remainder of the computational grid in one of the two following ways. 
\begin{itemize}
\item \textbf{Piecewise constant.} The parameter value at each point on the computational grid is set as the value of the nearest point on the parameter grid. 
\item \textbf{Piecewise linear.} The parameter value at each point on the computational grid is linearly interpolated from the two nearest points on the parameter grid. See Fig. \ref{approx}.
\end{itemize}

\begin{figure}[h]
	\centering
	\includegraphics[angle=0, width=1\textwidth]{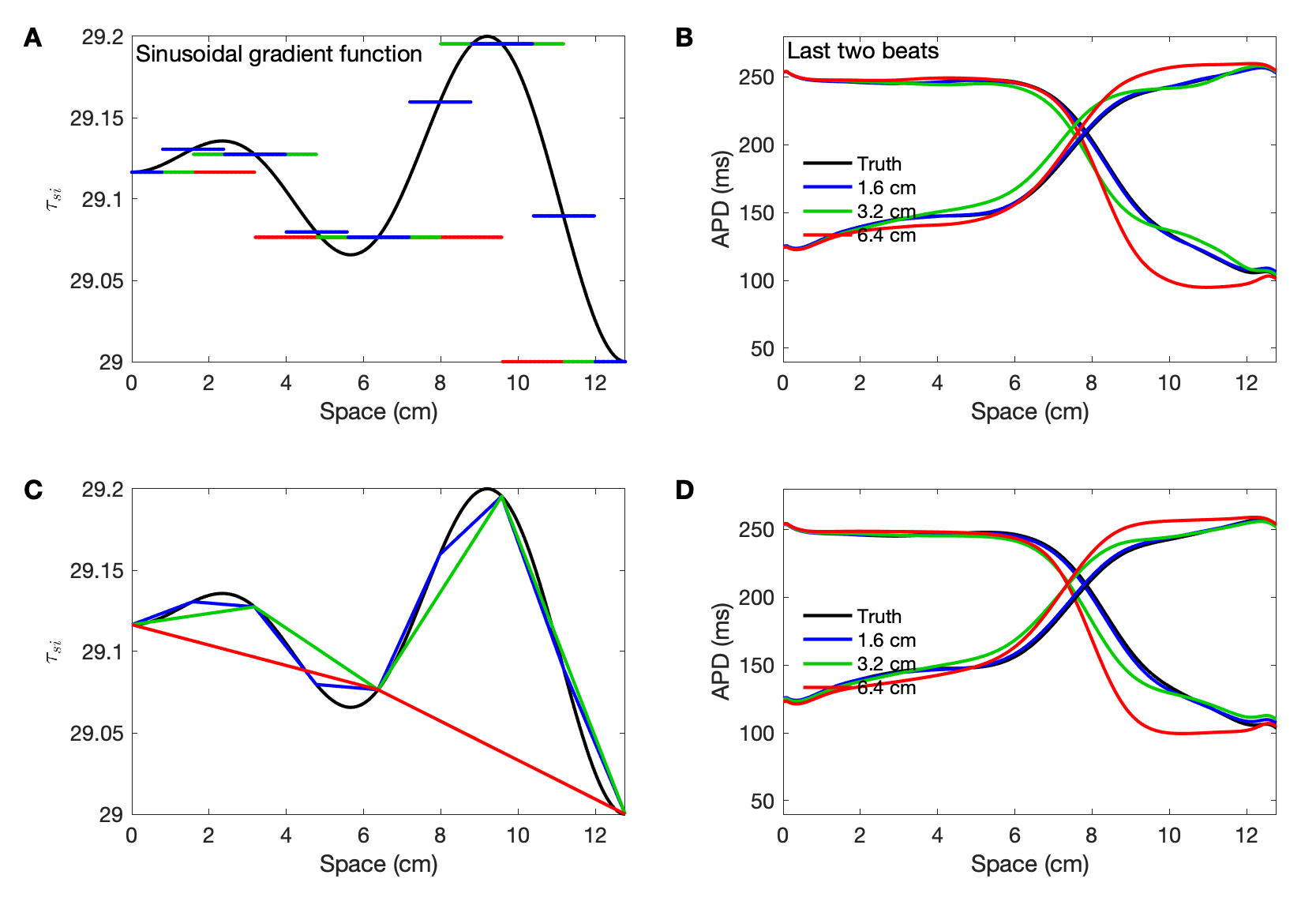}
	\caption{\label{alternans}Example of discordant alternans. These profiles were obtained for the $\tau_{si}$ parameter ---on a cable of 12.8 cm and using an original sinusoidal function (A,C) defined on the interval [29, 29.2]--- by pacing for 10 $s$ at a period of 325 ms to find the spatial profiles of action potential duration on the last two beats (B,D). The period was specifically selected to produce alternans and avoid block. A piecewise constant approximation was used in the first row (A,B) and a piecewise linear in the case of the lower row (C,D). This parameter resulted to be the more sensitive to changes on the range of the functions used.}
\end{figure}

Our goal is to identify the optimal spacing of the parameter grid to obtain good agreement with spatial profiles of action potential duration during complex states like discordant alternans; the idea behind this approach is that if our methodology is successful for complex states, even more so would  it be in the case of simpler states. Test parameter grid spacings were set logarithmically from a maximum spacing of 12.8 cm which represents half the length of the longer cable. See Fig. \ref{alternans}.

\begin{figure}[]
	\centering
	\includegraphics[angle=0, width=0.75\textwidth]{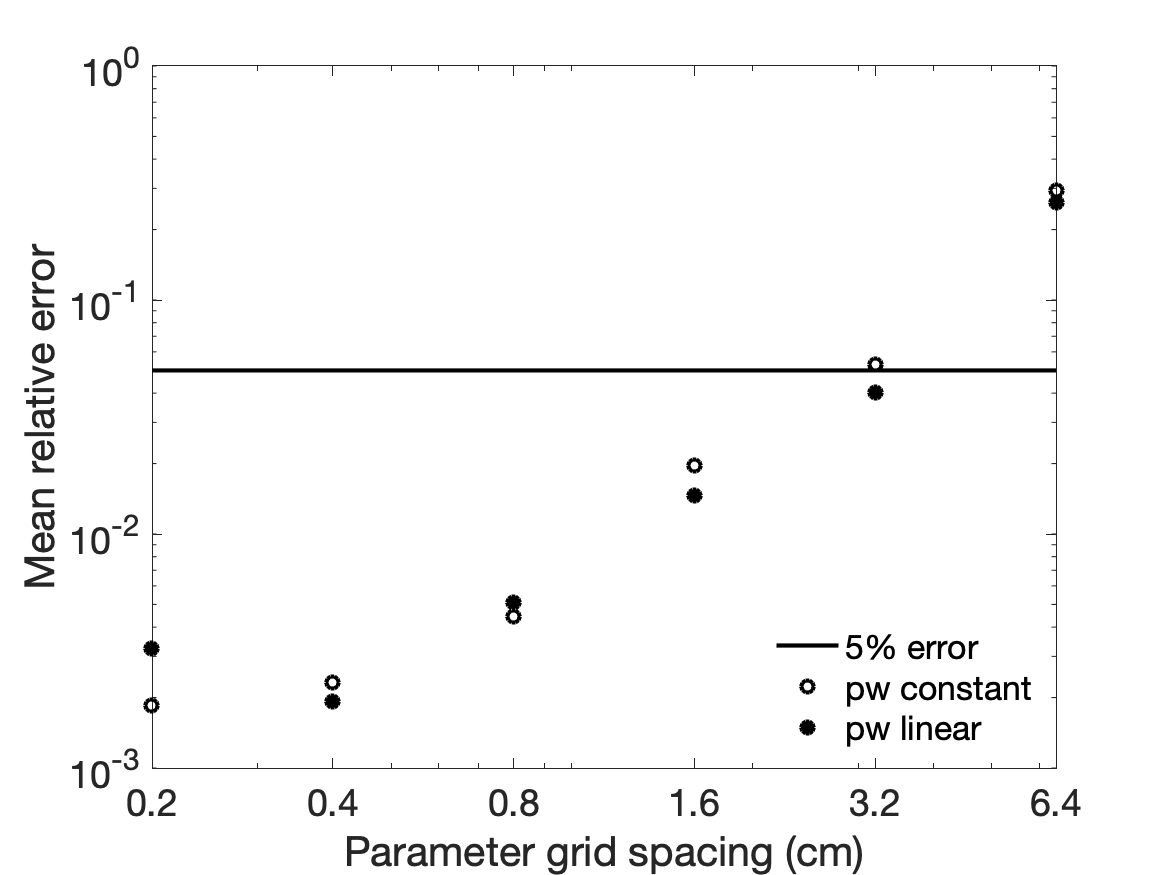}
	\caption{\label{error_example}Example of the average relative error over space for the parameter $\tau_d$ in the case of the cable of length 12.8 cm. The average was calculated over the eight functions used. Each node represents the approximation using different spacing to exemplify convergence: the finer the grid spacing the smaller the error. We selected a threshold of 5\% for the mean relative error.}
\end{figure}

To assess the accuracy of the approximations, we calculated the average relative error between the real and approximated action potential duration (APD) values. See Fig. \ref{error_example}.

\begin{figure}[]
	\centering
	\includegraphics[angle=0, width=1\textwidth]{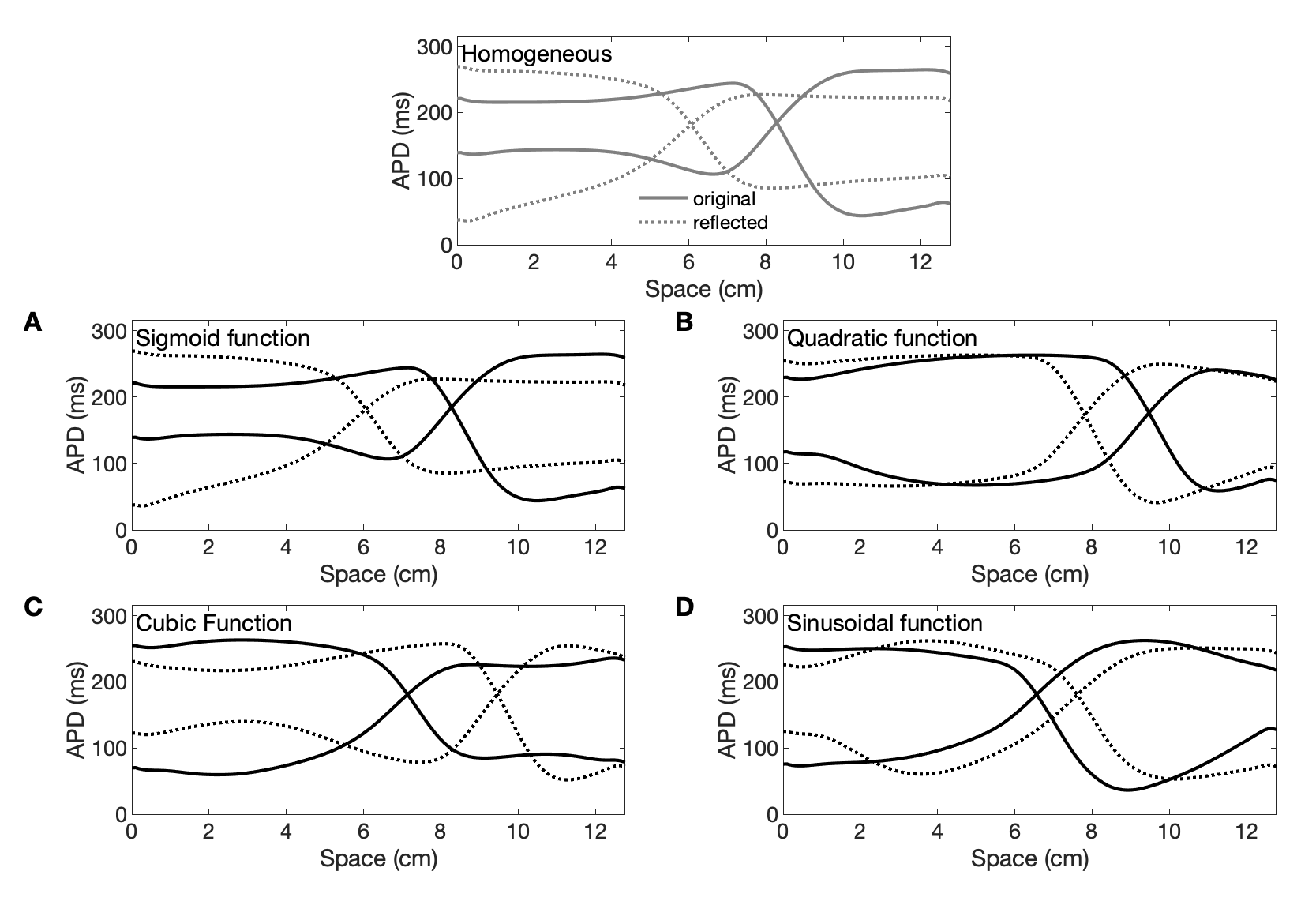}
	\caption{\label{homvshet}Example of different discordant alternans profiles for the parameter $\tau_r$ where homogeneity is compared to heterogeneity for a cable of length equal to 12.8 cm. For each figure, the full line graph represents the original function and the dotted line represents the reflected function.  On top, we see the graphs of the APD when homogeneity is assumed; the minimum value is 32.8 and the maximum is 33.4. Below, starting at the left upper corner, we see the profiles of the four gradient functions used when heterogeneity is considered: sigmoid (B), quadratic (C), cubic (D) and sinusoidal (E). The parameter value was varied taking values in the interval [32.8,33.4].
}
\end{figure}

In Fig. \ref{homvshet} we compare homogeneous to heterogeneous APD profiles to see the effect heterogeneity has in the model dynamics.

\section*{Examples}

In the following graphs, we present examples of spatial heterogeneity in the specified parameter using the different functions for several different parameter grid spacings. We show parameter values over the full domain (either of 12.8 or 25.6 cm) for both the piecewise linear and piecewise constant approaches for different parameter grid spacings. 
The average relative error over space is shown as a function of parameter grid spacing for a broader range of grid spacings to illustrate convergence; we selected an average relative error threshold of 5\%.

\begin{figure}[]
\begin{subfigure}{0.52\textwidth}
	\includegraphics[width=\textwidth]{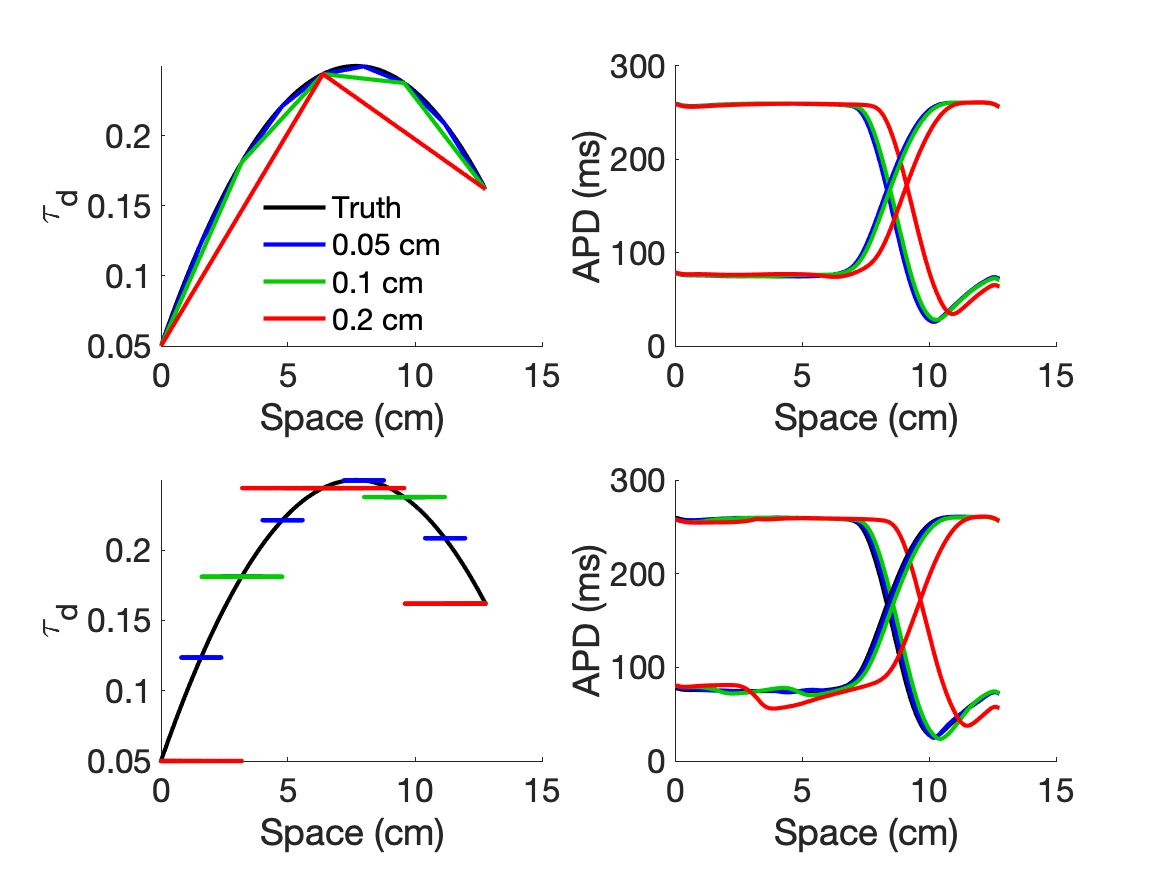}
	\caption{Short cable.}
\end{subfigure}
%
\begin{subfigure}{.52\textwidth}
	\includegraphics[width=\textwidth]{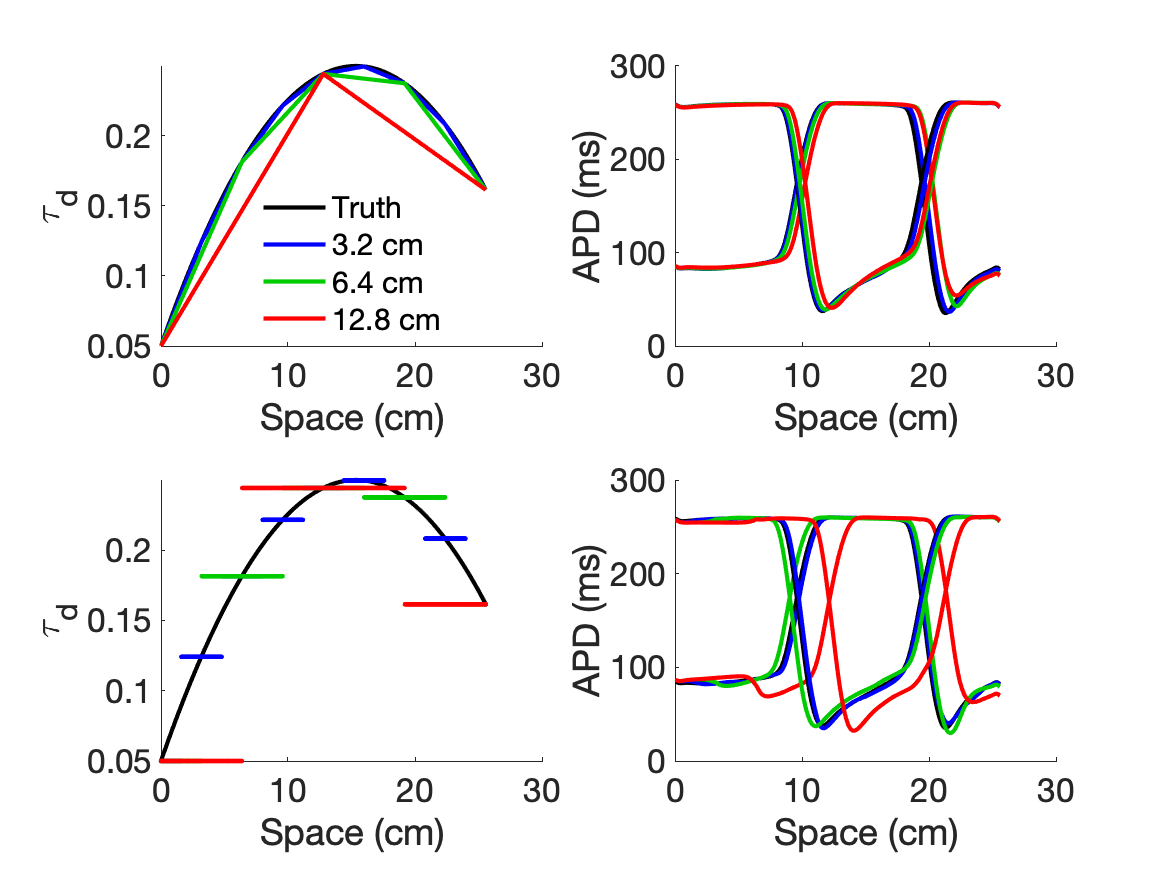}
	\caption{Long cable.}
\end{subfigure}
\caption{Parameter $\tau_d$, with an original quadratic gradient function and a maximum spacing of 12.8 (left) and 25.6 cm (right). In general, the number of nodes increased for the longe cable.}
\end{figure}

\begin{figure}[]
\begin{subfigure}{.52\textwidth}
	\includegraphics[width=\textwidth]{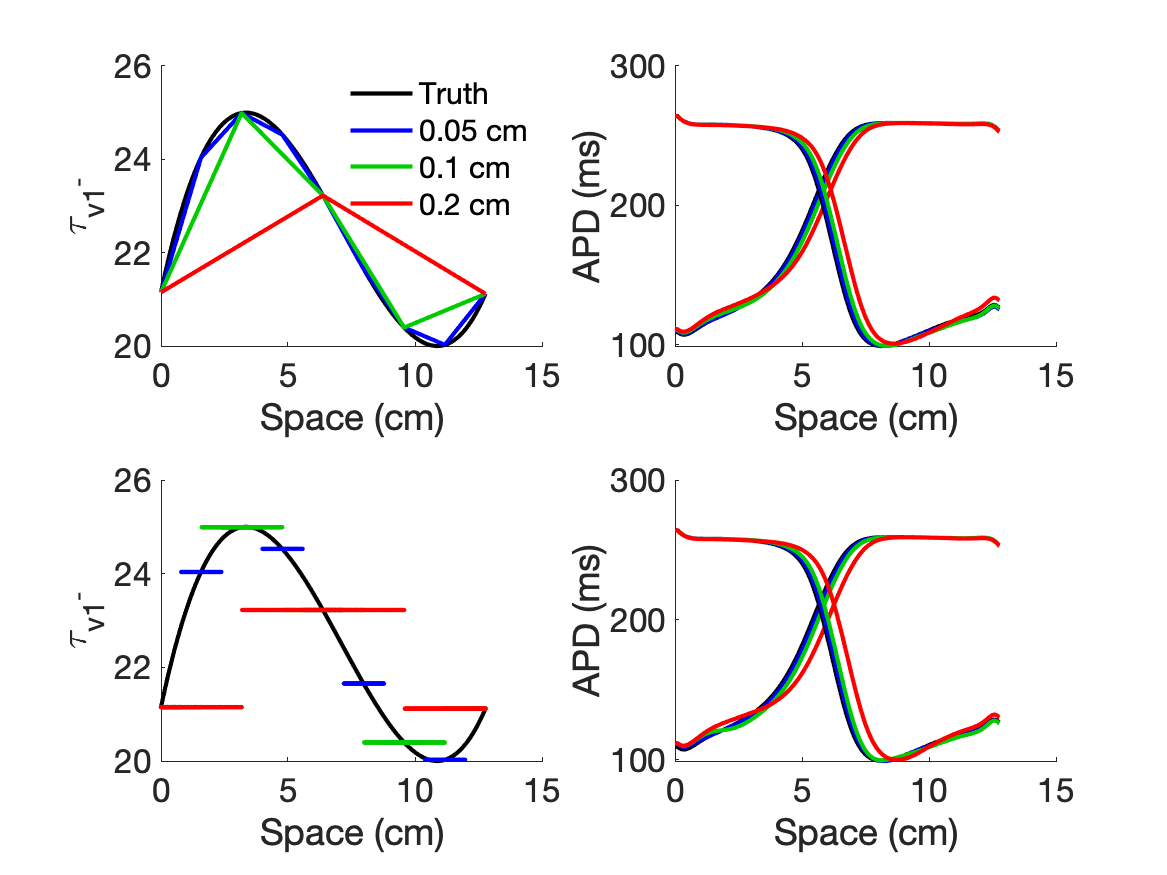}
	\caption{Piecewise approximations.}
\end{subfigure}
\begin{subfigure}{.52\textwidth}
	\includegraphics[width=\textwidth]{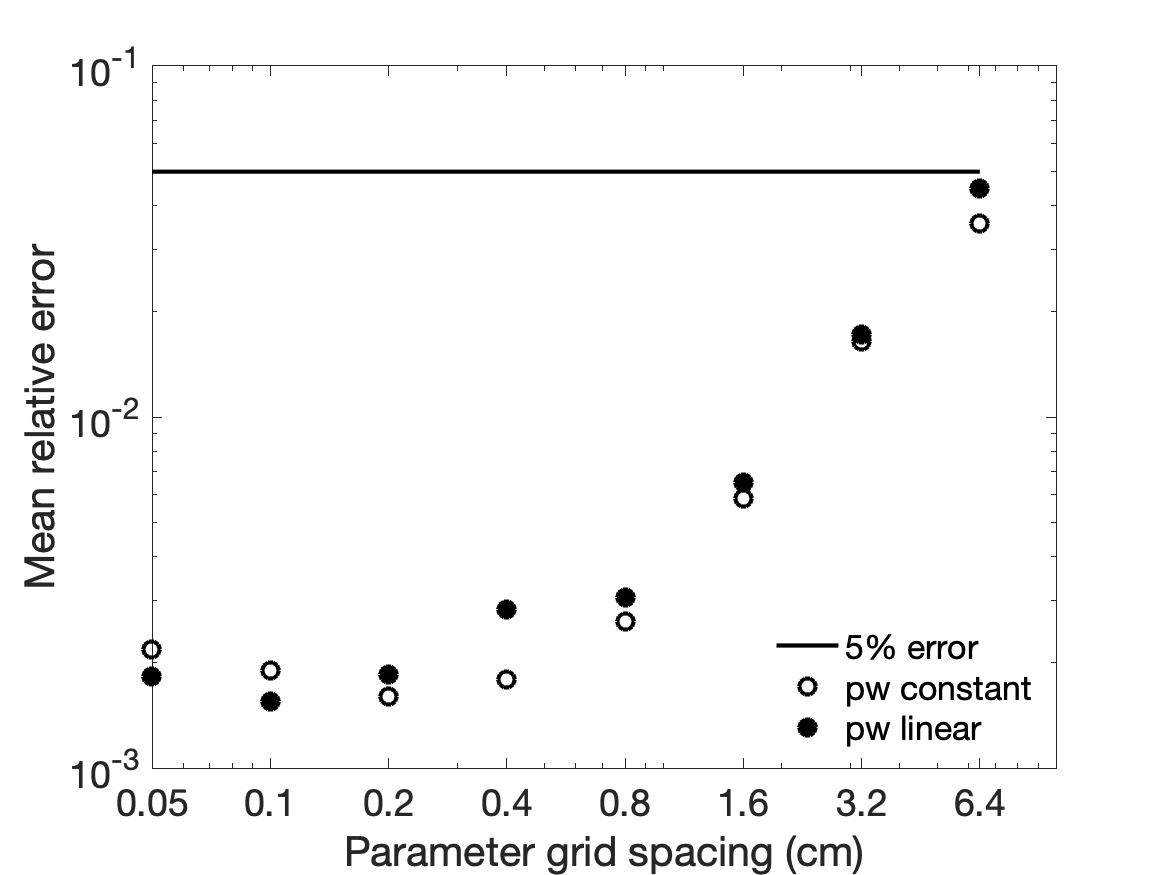}
	\caption{Relative error.}
\end{subfigure}
\caption{Parameter $\tau_{v1}^-$, with an original cubic gradient function and a maximum spacing of 12.8 cm.}
\end{figure}

\begin{figure}[]
\begin{subfigure}{.52\textwidth}
	\includegraphics[width=\textwidth]{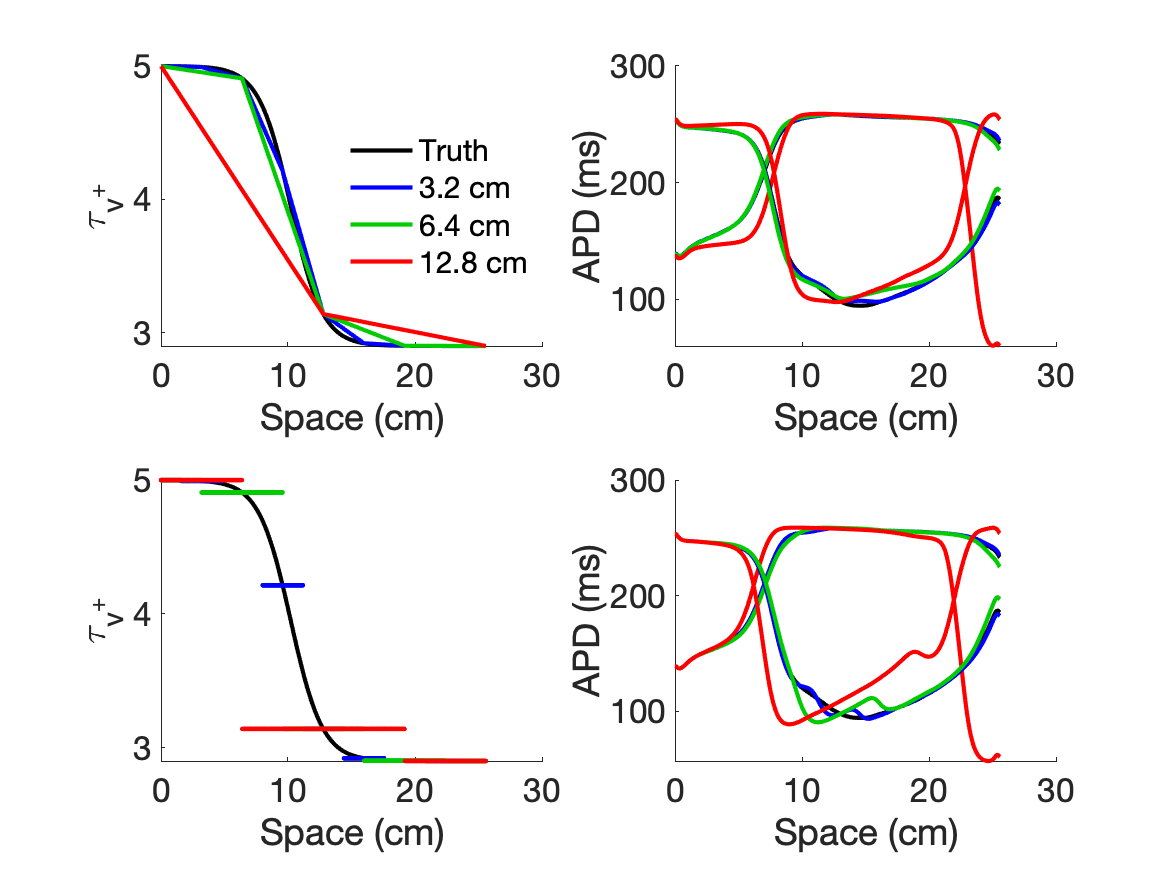}
	\caption{Piecewise approximations.}
\end{subfigure}
\begin{subfigure}{.52\textwidth}
	\includegraphics[width=\textwidth]{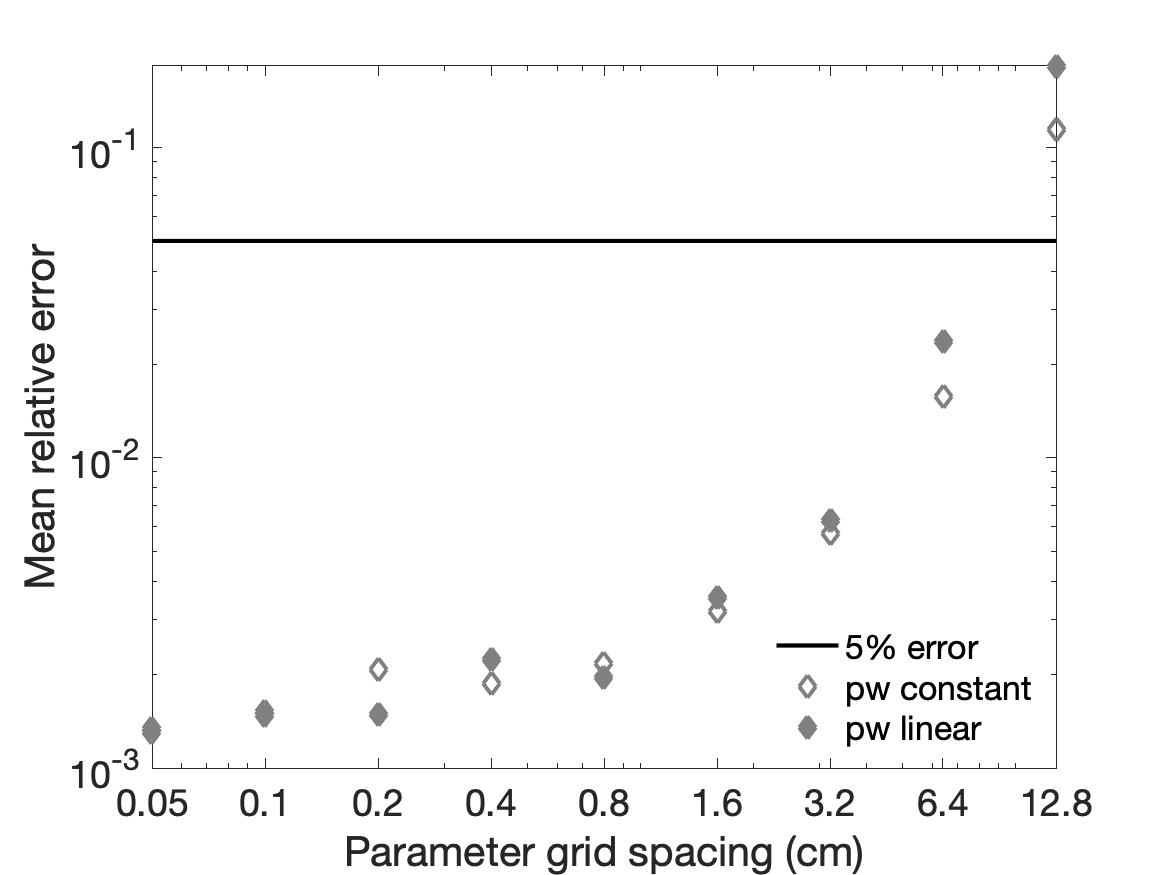}
	\caption{Relative error.}
\end{subfigure}
\caption{Parameter $\tau_{v}^+$, with a reflected sigmoid gradient function and a maximum spacing of 25.6 cm.}
\end{figure}

\begin{figure}[]
\begin{subfigure}{.52\textwidth}
	\includegraphics[width=\textwidth]{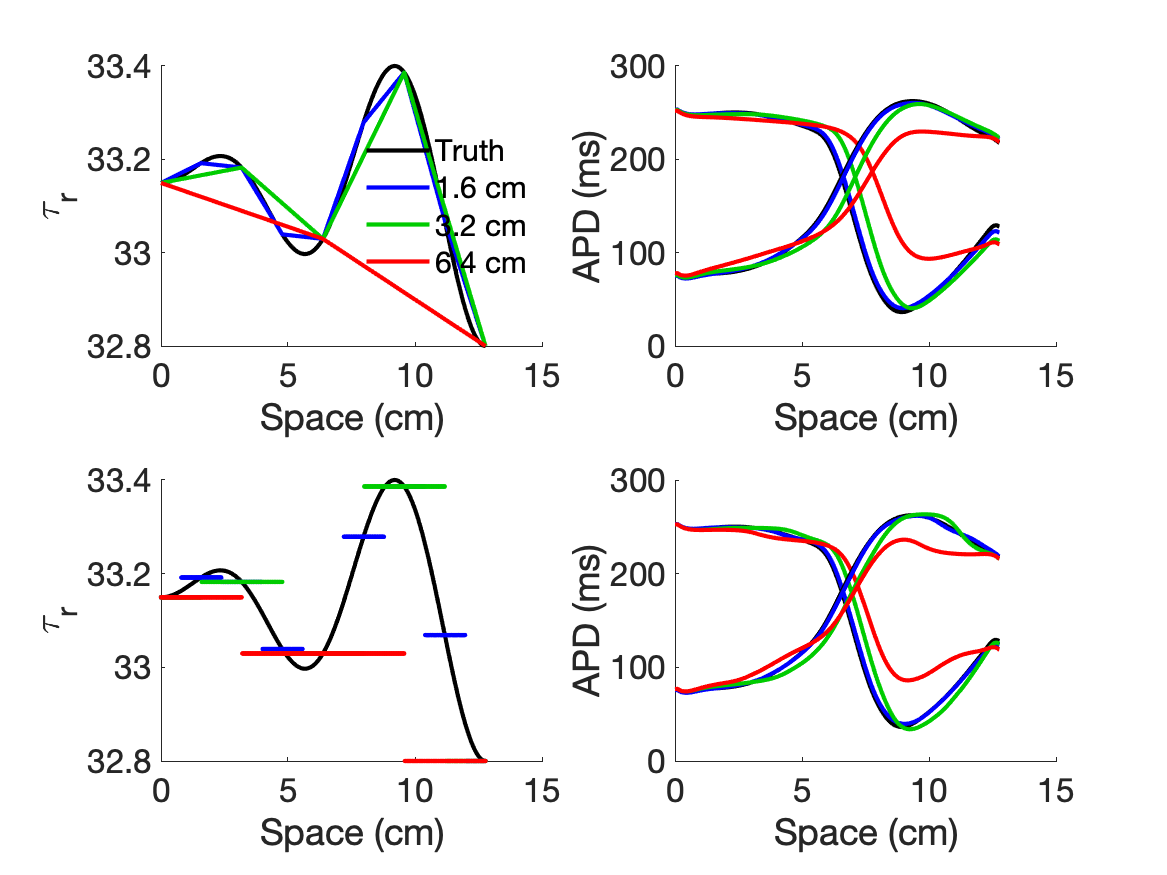}
	\caption{Piecewise approximations.}
\end{subfigure}
\begin{subfigure}{.52\textwidth}
	\includegraphics[width=\textwidth]{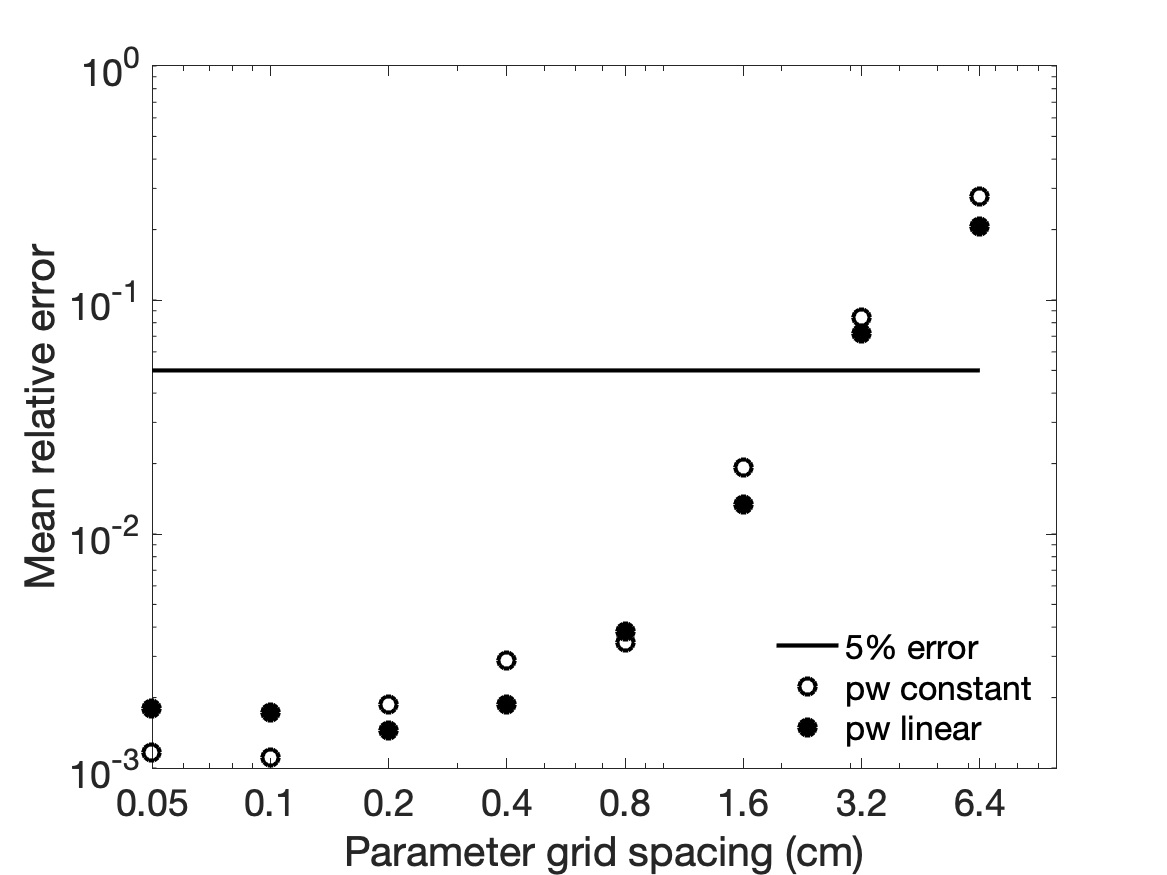}
	\caption{Relative error.}
\end{subfigure}
\caption{Parameter $\tau_{r}$, with a reflected sinusoidal gradient function and a maximum spacing of 12.8 cm.}
\end{figure}

\newpage
\section*{Results}

In general, for individual parameters, the relative error decreases as the grid spacing becomes finer, but at different rates. For some parameters such as $t_{v1}^-$, the relative error was below the threshold of 5\% for all grid spacings tested, considering both lengths.

Overall, for both cable lengths, we found that spacing of about 1.6 cm produces profiles within 5\% of the true profiles for a range of different model parameters and different functions of those parameters over space. In addition, the piecewise-linear approximations performed a little bit better than the piecewise-constant function approximations with one exception. In the case of the longer cable, it seemed that it was not possible to have an error below the threshold not even in the case of the finest grid used (0.05 cm) for the parameter $\tau_w^+$. The sinusoidal reflected gradient function causes this behavior. Taking a closer look, we realized that the reason of the error increasing to more than 5\% was due to the fact one of the alternans flipped.

 The maximum parameter grid spacing to achieve at most 5\% error across al functions can be seen on tables \ref{short} for the 12.8 cm length and \ref{long} for the 25.6 cm length. 

\begin{table}[]
\centering
\resizebox{\textwidth}{!}{\begin{tabular}{| c | c | c | c | c | c | c | c | c | c | c | c |}
\hline
Parameter & min. value & max. value & Period (ms) & PW Linear (cm) & PW Constant (cm)\\
\hline
$k$ & 9 & 10 & 300 & 3.2 & 3.2\\
$\tau_d$ & 0.05 & 0.25 &  305 & 3.2 & 1.6\\
$\tau_r$ & 32.8 & 33.4 & 305 & 1.6 & 1.6\\
$\tau_{si}$ & 29 & 29.2 & 325 & 3.2 & 3.2\\
$\tau_{v1}^-$ & 20 & 25 & 330 & 6.4 & 6.4\\
$\tau_v^+$ & 2.9 & 5 & 300 & 1.6 & 1.6 \\
$\tau_w^-$ & 35 & 110 & 300 & 3.2 & 1.6\\
$\tau_w^+$ & 750 & 870 & 300 & 3.2 & 1.6\\
\hline
\end{tabular}}
\caption{\label{short}For each parameter, the interval and cycle length used to produce alternans are shown. The maximum parameter grid spacing to achieve an average relative error of at most 5\% across all functions is presented for the 12.8 cm cable.}
\end{table} 

\begin{table}[]
\centering
\resizebox{\textwidth}{!}{\begin{tabular}{| c | c | c | c | c | c | c | c | c | c | c | c |}
\hline
Parameter & min. value & max. value & Period (ms) & PW Linear (cm) & PW Constant (cm)\\
\hline
$k$ & 9 & 10 & 325 & 6.4 & 6.4\\
$\tau_d$ & 0.05 & 0.25 &  305 & 6.4 & 3.2\\
$\tau_r$ & 32.8 & 33.4 & 310 & 3.2 & 3.2\\
$\tau_{si}$ & 29 & 29.2 & 325 & 6.4 & 6.4\\
$\tau_{v1}^-$ & 20 & 25 & 330 & 12.8 & 12.8\\
$\tau_v^+$ & 2.9 & 5 & 325 & 6.4 & 6.4 \\
$\tau_w^-$ & 35 & 110 & 335 & 6.4 & 3.2\\
$\tau_w^+$ & 750 & 870 & 305 & NA & 3.2\\
\hline
\end{tabular}}
\caption{\label{long}The interval and cycle length to produce alternans is presented for each parameter. The maximum parameter grid spacing to achieve an average relative error of at most 5\% across all functions is shown for the 25.6 cm cable.}
\end{table} 



\begin{figure}[h]
	\centering
	\includegraphics[angle=0, width=0.75\textwidth]{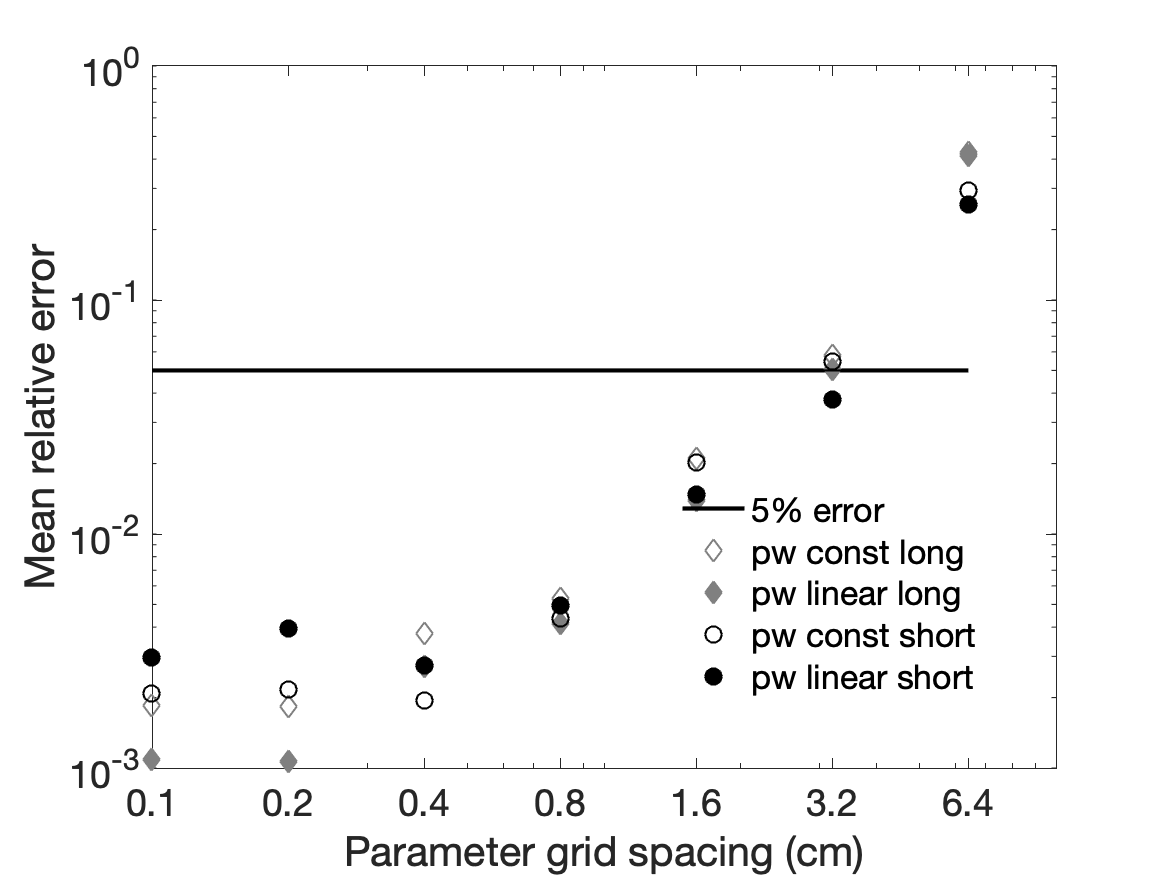}
	\caption{\label{errorcomparison} Comparison of the average relative error for the two different lengths (12.8 and 25.6 cm) used in this article for the parameter $\tau_d$. In general, we observed that considering a specific grid spacing, the error is a little lower when the cable is longer, independently of selecting the piecewise constant or piecewise linear approximation.}
\end{figure}

We compared the error for the two different lengths for each parameter and noted that overall, the error decreased for the longer cable. In Fig. \ref{errorcomparison} the graphs of the errors corresponding to the example presented in Fig. \ref{alternans} are shown. A special case occurred for the parameter $\tau_w^+$ where it was not possible to maintain the average relative error below 5\%.


	We mention that even though by selecting a finer grid reduces the error, this relation was not always monotonic. Because when calculating the action potential duration, we made approximations to get the voltage reaching a certain threshold, as we only had the values defined at certain grid points. Therefore, we directly compared the error of the voltage, but the behavior of the error was approximately the same as that one of the APD. 
	
	
	We also compared APDs corresponding to a large increase in errors when we expected them to decrease, but we did not find any particular pattern. We believe that the explanation to not having always a decreasing error as the grid becomes finer has to do with the propagation of electric waves on the cable by diffusion. For example, we produced profiles where the dynamics were simpler (without discordant alternans), and even though the results showed a more defined monotonic trend  in some cases, that did not happen in all cases.
		

				
The next step consisted in taking the signed error of the APD to see if we could detect any new behavior. As we did not see any new qualitative changes, we calculated the signed voltage of the difference between the real and the approximated value for all the parameters, getting approximately the same outcome as with the unsigned values. The error trend when using the APD or the voltage did not offer new information either, and for that reason, we sticked to making all the calculations of the errors with the APD instead of the voltage. 
	
We also focused on just one parameter and started with $k$. We found results that were not what we expected, like the fact that by varying this parameter with a gradient function has an effect on the wavefront. We noticed that the difference in magnitude of the fast inward current (between 0.6 and 0.7) is very large compared to the slow inward and outward currents (around 0.03).

Our results suggest that matching the output of models of cardiac tissue to heterogeneous experimental data can be done efficiently, even during complex dynamical states, and holds promise for more accurate modeling of individual experiments. 

\emph{This work is supported by National Science Foundation grant no. CNS-1446312.}

\end{document}